\documentclass{aa} 
\usepackage{natbib}
\usepackage{graphicx}
\usepackage{txfonts}
\usepackage{epsfig}
\usepackage{times}
\usepackage{rotating}
\usepackage{float}
\usepackage[caption = false]{subfig}
\usepackage{setspace}
\usepackage{lscape}
\usepackage{epstopdf}
\usepackage{tabularx}
\usepackage{caption}
\usepackage[usenames]{color}
\usepackage[toc,page]{appendix}
\usepackage{soul}
\usepackage{amsmath}

\begin{document}

\title{Galaxy transformation across the cosmic web: The influence zone of filaments}
\author{J. Alfonso L. Aguerri\inst{1,2} \& Stefano Zarattini\inst{3} }

\institute{Instituto de Astrof\'isica de Canarias, calle Vía L\'actea s/n, E-38205 La Laguna, Tenerife, Spain
\and Departamento de Astrof\'isica, Universidad de La Laguna, Avenida Astrof\'isico Francisco S\'anchez s/n, E-38206 La Laguna, Spain
\and Centro de Estudios de F\'isica del Cosmos de Arag\'on (CEFCA), Plaza San Juan 1, 44001 Teruel, Spain\\
jalfonso@iac.es, szarattini@cefca.es}

\date{\today}

\abstract{The matter distribution in the Universe exhibits a rich variety of structures, and they form the so-called cosmic web. These structures arise from the anisotropic gravitational collapse of primordial density fluctuations and define the pathways along which galaxies flow from low-density regions (voids) to the highest-density environments (galaxy clusters). The variation in local density across these structures plays a fundamental role in driving the environmental evolution of galaxies.}
{To characterise the boundaries of filaments, we derived and analysed the galaxy overdensity profiles around filaments in two redshift ranges: $0.05 < z < 0.1$ and $0.1 < z < 0.3$.}
{The profiles perpendicular and parallel to the filament directions were derived by averaging the galaxy overdensity as a function of distance. The characteristic scales and central overdensities of these profiles were then analysed by fitting several analytical models, grouped into two main families: exponential and power-law models. We also introduced normalised density profiles to analyse the impact of survey incompleteness.}
{ The galaxy overdensity profiles in the perpendicular direction to filaments show an almost constant value in the central regions $D_{fila} < 1$ Mpc, with a decrease at larger distances until $\approx 10$ Mpc. The mean physical widths of the filaments at redshifts $0.05<z<0.1$ and $0.1<z<0.3$, measured as the mean scale radii, are $2.39\pm0.69$ and $5.56\pm2.29$ Mpc, respectively.  The difference in the scales in the two redshift ranges is also observed in the normalised density profiles. On the other hand, density profiles along filaments show a constant behaviour for distances larger than about 20 Mpc from the nearest intersection.}
{The results presented in this work show that the zone of influence of cosmic filaments extends up to $\sim$10 Mpc from their spines. Moreover, a mild evolution in filament structural parameters is observed over the past $\sim$4 Gyr, suggesting that filaments undergo measurable changes even at relatively low redshifts.}

\keywords{large-scale structure of Universe}

\authorrunning{J. A. L. Aguerri \& S. Zarattini}
\titlerunning{The influence zone of filaments}
\maketitle

\section{Introduction}
\label{sec:intro}

The matter distribution in the Universe, on megaparsec scales, is not homogeneous. Instead, it exhibits a rich variety of structures that form the so-called large-scale structure of the Universe. These structures are the consequence of gravitational evolution and the anisotropic collapse of the initial fluctuations in the primordial density field \citep{Zeldovich1970}. 

The large-scale structure of the Universe forms the network of roads and highways along which dark matter (DM) flows, moving from low-density regions (cosmic voids) through intermediate-density structures (walls and filaments), and eventually accumulating in the highest-density regions (nodes). This continuous flow of matter shapes the skeleton of the cosmic web \citep{Bond1996, AragonCalvo2010, Cautun2014}. Baryonic matter subsequently falls into these DM haloes and cosmic structures due to gravity, leading to the formation of galaxies and clusters \citep{White1978, Dekel1986, Springel2005}.

Pioneering observations of the galaxy distribution around the Perseus cluster revealed that cluster locations are not random. Instead, they lie along chains of clusters that form even larger gravitationally bounded structures, known as superclusters \citep{Joeveer1978}. This picture was later confirmed by the first galaxy redshift surveys, such as the Center for Astrophysics (CfA) galaxy survey \citep{deLapparent1986}. 

Today, the large-scale structure of the Universe is mapped and analysed using data from extensive spectroscopic surveys that cover large portions of the sky with unprecedented systematic precision. Notable examples include the Sloan Digital Sky Survey \citep[SDSS;][]{York2000}, the Two Degree Field Galaxy Redshift Survey \citep[2dFGRS;][]{Colless2001}, the 6dF Galaxy Survey \citep[6dFGS;][]{Jones2009}, the Galaxy and Mass Assembly survey \citep[GAMA;][]{Driver2009}, and the WISExSuperCOSMOS survey \citep[WISExSCOS;][]{Bilicki2016}.

Theory and numerical simulations have been crucial for understanding the formation of structure in the Universe. The theoretical framework describing how the initial fluctuations grew into the filamentary structure observed today was first developed by \citet{Zeldovich1970}. That work laid the foundation for subsequent studies based on numerical simulations. One of the earliest numerical simulations using cold dark matter (CDM) particles was carried out by \citet{Davis1985}, who demonstrated that the large-scale structures observed in the Universe naturally arise from the initial perturbations predicted by the CDM model. More recently, ambitious large-scale simulations such as the Millennium Simulation \citep{Springel2005}, and the Illustris and IllustrisTNG projects \citep{Vogelsberger2014, Springel2018}, have enabled detailed studies of clusters, filaments, galaxy haloes, and their evolution within the cosmic web. These simulations have shown that at $z = 0$, voids occupy approximately $\sim 77\%$ of the volume of the Universe, walls around $\sim 18\%$, filaments $\sim 6\%$, and nodes less than $0.1\%$. In contrast, the distribution of mass is markedly different: About $50\%$ of the total mass resides in filaments, $24\%$ in walls, $15\%$ in voids, and $11\%$ in nodes \citep[see][]{Cautun2014}. These predictions are broadly consistent with recent observational studies that attempt to classify the cosmic web using galaxy positions and redshifts from large spectroscopic surveys \citep[e.g.][]{Tempel2014, Laigle2018}, although differences in classification methods and projection effects can lead to discrepancies in the estimated fractions.

The impact of dense environments, such as galaxy clusters, and underdense regions, such as voids, on galaxy properties has been extensively studied in the literature. Observations have shown that galaxy clusters host predominantly early-type, red, and passive galaxies \citep{Dressler1980, Balogh2004, Boselli2006, Aguerri2007, RomeroGomez2024a}, which are affected by processes such as ram-pressure stripping \citep{Gunn1972, Quilis2000}, harassment \citep{Moore1996, Moore1998}, and strangulation \citep{Bekki2002, Fujita2004}. In contrast, galaxies in voids tend to be star forming, blue, and late-type, reflecting a slower evolutionary path due to the lower density of their environment and fewer interactions with other galaxies \citep{Rojas2004, Hoyle2005, Kreckel2012}. These environmental dependencies highlight the crucial role of the environment in shaping galaxy evolution. There is growing evidence that significant transformations in galaxy properties begin before galaxies enter clusters, driven by various physical processes collectively referred to as pre-processing \citep[][]{Sarron2019}.

In contrast, the impact of filaments and walls on galaxy properties is still not well understood. This limited understanding may be due to the fact that 
extensive, reliable, and precise filament catalogues in the nearby Universe only became available in the past decade \citep[see][]{Chen2016, Li2016, Malavasi2017, CarronDuque2022}. These catalogues typically provide detailed information on the spatial distribution of filaments, often derived from galaxy redshift surveys. They identify the locations and morphologies of large-scale structures, such as filaments and walls, by analyzing the spatial correlations of galaxies and using algorithms such as the persistence diagram or the density field reconstruction. The filaments are typically defined as elongated structures connecting galaxy clusters, and their properties—such as length, orientation, and density—are catalogued alongside the galaxy population residing in or near them. The catalogues are crucial for understanding how environmental factors associated with the cosmic web affect galaxy evolution, as they link galaxy properties to their positions in these large-scale structures.

 Increasing evidence suggests that galaxy evolution and quenching do not occur exclusively within galaxy clusters. On the contrary, changes in galaxy properties already begin to emerge at the level of filaments.  In particular, the fraction of red galaxies, for a given stellar mass, turned out to be higher in filaments than in the field  \citep[][]{Kraljic2018, Laigle2018, Pandey2020, Hoosain2024}, which contributes to galaxies displaying redder stellar colours and lower star formation rates near filaments \citep[][]{Rojas2004, Kuutma2017, Luber2019, Zarattini2025}. In addition,  \cite{Kuutma2017} found that the morphology of galaxies also changes with the distance to filaments. Thus, they found that the elliptical-to-spiral ratio varies with the distance to filaments, becoming larger at distances closer to filaments. \cite{Chen2017} discovered that galaxies closer to filaments are redder, larger, and more massive than those located at large distances. These galaxy properties observed in the nearby Universe are similar at higher redshift. \citet{Malavasi2017} found that the most massive and quiescent galaxies are closer to the filaments in the VIPERS survey at $z \sim 0.7$.

Observational studies analyzing the density profiles of filaments and their zones of influence are still limited, but their number is steadily growing. Notable exceptions include the work by \cite{Castignani2021} and \cite{Castignani2022a, Castignani2022b}, who analysed the density profile of a sample of 13 filaments surrounding the Virgo cluster. Their results suggest that filaments span scales of a few megaparsecs and that their physical properties evolve with redshift, with indications that the cosmic web's filamentary structure becomes more defined at lower redshifts. The identification and study of these filaments are crucial for understanding how environment-driven processes in the cosmic web affect galaxy properties such as star formation rates, morphology, and gas content.

The purpose of the present work is to analyse the properties of filament density profiles in a sample of filaments that covers a redshift range of $0.05 < z < 0.3$. We aim to trace the influence zone of filaments where galaxies can evolve due to the presence of these cosmic structures. The filament sample was obtained from the SDSS and was measured by \cite{Chen2016}. In a recent study, \cite{Zarattini2025} showed that stellar colours and star formation rates exhibit distinct transformations at different distances from the filaments. The characterisation of the filament density in this work provides valuable insights into the observed galaxy transformations. Specifically, by comparing the galaxy properties with the density profiles of the surrounding filaments, we could explore how the filamentary structures influence the environment of galaxies and trigger processes such as quenching, morphological evolution, and variations in star formation activity. This analysis will not only enhance our understanding of the role of filaments in galaxy evolution but also shed light on how the large-scale cosmic environment, as traced by the cosmic web, contributes to the transformation of galaxies over cosmic time.

The paper is organised as follows. The filament sample is described in Section~2. The filament overdensity profiles are presented in Section~3. In Section~4, we analyse the typical scales and the central galaxy overdensities within the filaments. The discussion and conclusions are provided in Sections~5 and~6, respectively. Throughout this paper, we adopt a flat $\Lambda$CDM cosmology with $\Omega_{m} = 0.3$, $\Omega_{\Lambda} = 0.7$, and $H_{0} = 70$ km s$^{-1}$ Mpc$^{-1}$.

\section{Sample of galaxies and their cosmic web location}
\label{sec:sample}

The data used in the present work comes from \citet{Zarattini2025}. They made a compilation of data from SDSS Data Release 16 \citep[DR16,][]{Ahumada2020} and used the catalogue from \citet{Chen2016} to trace the large-scale structure. The data obtained from the SDSS-DR16 correspond to all objects catalogued as galaxies brighter than $r_p =17.77$, the so-called main spectroscopic sample, where $r_p$ represents the $r-$band Petrosian magnitude. In addition, the final galaxy classification was filtered by using their position in the $r \ - <\mu(r_{50})>$ plane \citep[see e.g. ][]{Sanchez-Janssen2005,Aguerri2020}. Here, $<\mu(r_{50})>$ represents the mean surface brightness of the object within its effective radius ($r_{50}$). This filtering reduces the number of misclassified stars in the galaxy sample.

\begin{figure}
    \centering
    \includegraphics[width=0.5\textwidth]{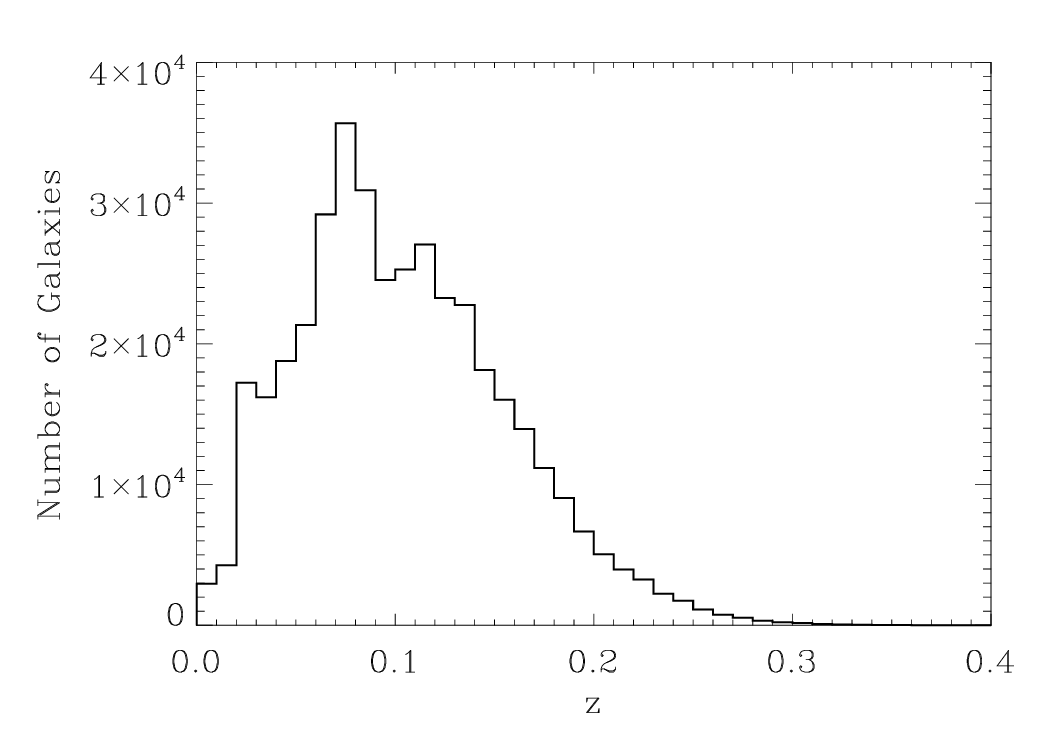}
    \caption{Redshift distribution of the galaxies used in this work.}
    \label{fig:z_dist}
\end{figure}

The final selected galaxies span a large area in the sky with $130^{o} < RA (J2000) < 230^{o}$ and $0^{o} < DEC (J2000) < 60^{o}$, a region where the cosmic web catalogue is uniformly mapped in \citet{Chen2016}. This final sample contains a total of $394\,116$ galaxies spanning a redshift range $0<z<0.69$. However, the redshift distribution varies considerably across the range.  Figure \ref{fig:z_dist} shows the redshift distribution of the galaxies used in the present work. Note that the galaxy counts peak at $z \approx 0.07$ and decline at higher redshifts.

The number of galaxies in the redshift intervals $z<0.1$, $0.1<z<0.2$, $0.2<z<0.3$, and $z>0.3$ are $201\,104$ (51.0$\%$), $173\,358$ (43.9$\%$), $19\,189$ ($4.9\%$) and 465 ($0.1\%$), respectively. In this work, we focus on the properties of filaments in the redshift range $0.05 < z < 0.3$, which contains more than 85$\%$ of the $394\,116$ galaxies in our sample. The remaining 15$\%$ are mainly located at $z < 0.05$, a range where the cosmic web catalogue of \citet{Chen2016} is not available.

\section{Filaments overdensity profiles}
\label{sec:filaments}

\citet{Zarattini2025} measured the projected local galaxy density ($\Sigma_{g}$) around each galaxy in the SDSS sample. This density was calculated by identifying the five nearest neighbours within a velocity range of $\pm 3000$ km s$^{-1}$. To account for spectroscopic incompleteness, these authors applied a photometric correction to $\Sigma_{g}$. The local galaxy overdensity was then defined as $\delta_{g} = (\Sigma_{g} - \Sigma_{z,g})/\Sigma_{z,g}$, where $\Sigma_{z,g}$ is the mean galaxy density within a redshift slice of $\Delta z = 0.005$. This normalisation minimises redshift-dependent biases in the overdensity measurement. \citet{Zarattini2025} also computed the minimum projected distance from each galaxy in the SDSS sample to the nearest filament ($D_{fila}$) and to the nearest intersection ($D_{int}$), using the filament catalogue from \citet{Chen2016}. The Cosmic Web was divided by \citet{Chen2016} into redshift slices of $\Delta z = 0.005$. Within each slice, filaments are represented as a series of points in right ascension and declination. All the points in the same slice are assumed to be at the same redshift and represent a two-dimensional map of the cosmic web. Although these points are not contiguous, their density is sufficient to trace the filament spine. Galaxies are also identified by their right ascension and declination. Assuming that galaxies and filaments lie at the same redshift (i.e., within the same redshift slice), we computed $D_{\mathrm{fila}}$ as the minimum projected angular distance between each galaxy and the nearest filament spine. This angular distance was then converted into a physical distance (in Mpc) using the angular  distance at the corresponding redshift, according to our adopted cosmology.

\subsection{Radial filament galaxy overdensity profiles}

\begin{figure*}
    \sidecaption
    \includegraphics[width=12cm]{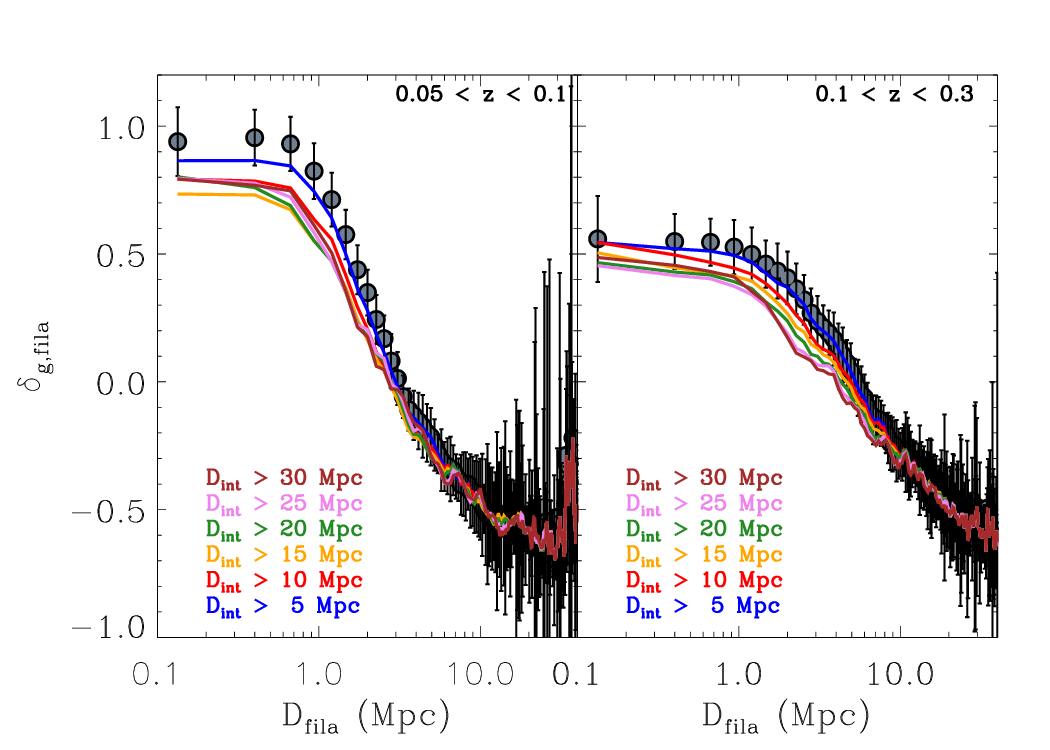}
    \caption{Galaxy overdensity as a function of radial distance from the filament spine for galaxies at $0.05 < z < 0.1$ (left panel) and $0.1 < z <0.3$ (right panel). Different symbols and colours indicate the overdensity profiles obtained by excluding galaxies within various distances from the nearest filament intersection: $D_{int} > 0$ (grey dots), $D_{int} > 5$ Mpc (blue line), $D_{int} > 10$ Mpc (red line), $D_{int} > 15$ Mpc (yellow line), $D_{int} > 20$ Mpc (green line), $D_{int} > 25$ Mpc (pink line), and $D_{int} > 30$ Mpc (brown line). Uncertainties are calculated as $\sigma/\sqrt{N_{p}}$, where $\sigma$ is the standard deviation and $N_{p}$ is the number of galaxies in each radial bin.        \\
    \\
    \\
    \\}
    \label{fig:overdens_perp}
\end{figure*}

To measure the radial galaxy overdensity profiles around filaments ($\delta_{g,fila}$), we computed the median galaxy overdensity in radial bins from the filament spine within two redshift ranges $0.05 < z < 0.1$ and $0.1 < z < 0.3$. Figure \ref{fig:overdens_perp} shows the resulting radial overdensity profiles (grey points). These profiles were derived using a total of $93\,307$ galaxies located within 40 Mpc of the filament spines. The computation of the profiles were done in uniformly spaced bins of projected distance $D_{\mathrm{fila}}$. We adopted a uniform binning in $D_{fila}$ in order to preserve a constant physical scale and to allow a direct, point-by-point comparison between different galaxy subsamples. Although the overdensity becomes asymptotic beyond $\sim 10$ Mpc, the outer bins are shown only to illustrate the convergence to the background level. Using variable-width bins at large $D_{fila}$ would not change the inferred asymptotic behaviour nor improve the fit constraints, which are intrinsically dominated by the inner regions of the profile. The median value of the overdensity in each bin is shown to mitigate the effect of potential outliers or skewness in the $\delta_{\mathrm{fila}}$ distribution. The uncertainties shown in Fig. \ref{fig:overdens_perp} were obtained as $\sigma/\sqrt{N_{p}}$, where $\sigma$ is the standard deviation of the galaxy overdensity and $N_{p}$ the number of galaxies in each radial bin, which ranges from over 1000 in the central regions ($D_{\mathrm{fila}} < 9$ Mpc) to fewer than 100 at large distances ($D_{\mathrm{fila}} > 21$ Mpc), with some bins containing as few as 6 galaxies. This naturally leads to larger error bars and increased scatter at large $D_{\mathrm{fila}}$. When fitting analytical models to the profiles, each data point is weighted by its uncertainty. As a result, the fits are primarily constrained by the inner regions of the profile, where the number of galaxies per bin is highest and the uncertainties are smaller.

In both redshift bins, the overdensity $\delta_{g,fila}$ remains approximately constant for distances $D_{fila} < 1$ Mpc, and then gradually declines, reaching an asymptotic value at around 10 Mpc. The drop in overdensity from the filament centre to 10 Mpc is approximately 1.5 and 1.0 for the profiles of the filaments at $0.05 < z < 0.1$ and $0.1 < z < 0.3$, respectively, which is more than three times the uncertainty at that distance. Notice that filaments at $0.1 < z < 0.3$ display profiles with a smaller central galaxy overdensity and a shallower decline compared to those at a lower redshift.

Similar profiles of $\delta_{g,fila}$ have been reported by other authors in the literature. In particular, our results are consistent with those reported by \citet{Tanimura2020}, \citet{Bonjean2020}, \citet{Castignani2022a, Castignani2022b}, and \citet{Wang2024}. Filament density profiles have also been studied in cosmological simulations. For example, \citet{Galarragaespinosa2020} found comparable profiles when analyzing filaments in the IllustrisTNG simulations. This qualitative agreement between observations and simulations supports the robustness of the measured filament density profiles.

We also analysed how the exclusion of galaxies located at different distances from intersections ($D_{int}$) impacts on the $\delta_{g,fila}$ profile. Figure \ref{fig:overdens_perp} also shows the $\delta_{g,fila}$ profiles after masking galaxies at various $D_{int}$ thresholds. In all cases, the overall shape of the profiles is similar regardless of the masking radius. In contrast, the overdensity values of the profiles decrease as the masking radius increases, which is expected since the highest galaxy overdensities are found near intersections.  However, the profiles remain similar when galaxies at $D_{int} \geq 10$ Mpc are included. Based on this result, we adopted a masking threshold of $D_{int} = 10$ Mpc for computing the radial $\delta_{g,fila}$ profiles for the two redshift bins.

\citet{Chen2016} identified intersections as crossing points of galaxy filaments. However, this should not be directly interpreted as evidence that these intersections correspond to the positions of galaxy clusters, as might be expected from the conventional view of the cosmic web. In fact, in \citet{Zarattini2025} it was shown that the overdensity of galaxies in intersections is about twice that of filaments and far from the typical overdensities of galaxy clusters. The variation that we observe in the central regions of the density profiles reflects this small increase in local overdensity, rather than indicating the presence of galaxy clusters. However, it is clear that intersections represent the highest density regions in the \citet{Chen2016} catalogue.

\subsection{Galaxy overdensity profile along filaments}

\begin{figure*}
    \sidecaption
    \includegraphics[width=12cm]{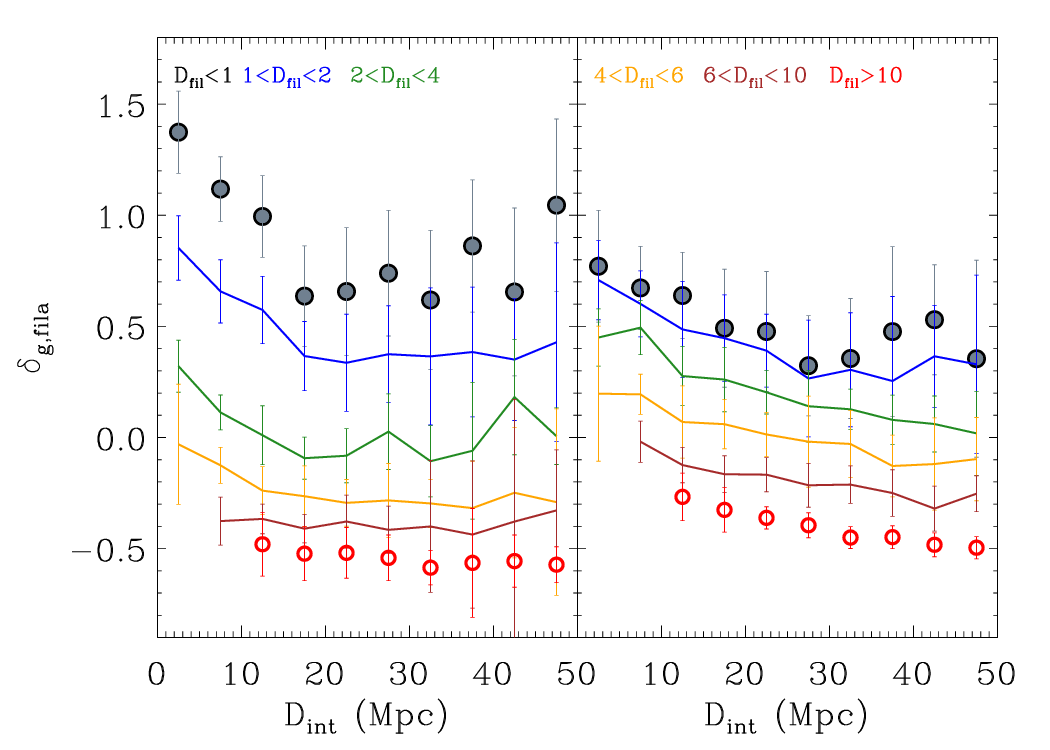}
    \caption{Galaxy overdensity profile along filaments for galaxies at $0.05 < z < 0.1$ (left panel) and $0.1 < z <0.3$ (right panel). The different symbols represent profiles taking into account galaxies at different distances from the filament spine: $D_{fila} < 1$ Mpc (grey points), $1<D_{fila}<2$ Mpc (blue line), $2<D_{fila}<4$ Mpc (green line), $4<D_{fila}<6$ Mpc (yellow line), $6<D_{fila}<10$ Mpc (brown line), and $D_{fila} > 10$ Mpc (red points). Uncertainties are similar to Fig \ref{fig:overdens_perp}. 
    \\
    \\
    \\
    \\}
    \label{fig:overdens_para}
\end{figure*}

Figure \ref{fig:overdens_para} shows the median galaxy overdensity values along filaments, i.e., as a function of the distance from the intersections for the two redshift bins. This figure presents the galaxy overdensity at the filament spine (grey points, for galaxies with $D_{fila} < 1$ Mpc) as well as the overdensities considering galaxies at varying distances from the filament spine  (coloured lines). It is important to note that in both redshift ranges the galaxy overdensity is the highest at the filament spine and remains constant for distances from the intersections larger than approximately 20 Mpc. At smaller distances, the overdensity increases due to the influence of the intersections. For galaxies in the redshift bin $0.05 < z < 0.1$, this behaviour is consistent across profiles up to $D_{fila} < 6$ Mpc. For larger distances from the filament spine, the overdensity profiles are consistently flat (see Fig. \ref{fig:overdens_para}). Note that the highest galaxy overdensity in Fig.~\ref{fig:overdens_para} is larger than in Fig.~\ref{fig:overdens_perp}. This is expected, as the profile in Fig. \ref{fig:overdens_para} always starts from the centre of intersections, where the overdensity is higher, and is an average along the entire filament range.

These kinds of profiles have not been extensively analysed in either observations or simulations. However, \cite{Castignani2022a} studied the galaxy density profiles around filaments in the Virgo cluster, finding that the filament density decreases more gradually with distance from the cluster, with a typical scale of $13.84 \pm 4.49$ Mpc (in our adopted cosmology), which corresponds to approximately four times the filament width in the perpendicular direction. In addition, \cite{Cautun2014}, using DM only  numerical simulations, found filament density profiles along the filament direction that are comparable to those presented in this work (see their Fig.~15). 

Figure~\ref{fig:overdens_para} also reveals a clear gradient in galaxy overdensity as a function of distance from the filament spine. In particular, the local galaxy density at the core of the filaments ($D_{\mathrm{fila}} < 1$~Mpc) is approximately 2.5 times higher than at $D_{\mathrm{fila}} \approx 5$~Mpc for filaments at $0.05<z<0.1$. This density contrast is reduced to about 1.5 times for filaments at $0.1<z<0.3$. This density gradient suggests that galaxies may undergo significant environmental transformations as they migrate towards the denser regions of the filaments.

Similar gradients in galaxy density and properties with filament proximity have been reported in previous studies. For example, \citet{Malavasi2017} observed a clear modulation of galaxy star formation activity with distance to the filament, while \citet{Kraljic2018} and \citet{Laigle2018} also reported systematic variations in galaxy morphology and star formation rates as a function of filament proximity using the COSMOS2015 and Horizon-AGN simulations, respectively. These works support the idea that the physical conditions within and around filaments are a function of the filament distance and could play a key role in driving galaxy evolution.

Filaments in the redshift range $0.1 < z < 0.3$ appear broader and less overdense compared to those at $0.05 < z < 0.1$. This is evident in Fig.~\ref{fig:overdens_para}, where both the central overdensity at the filament spine and the gradient of the overdensity profile---measured between the spine and regions at $D_{\mathrm{fila}} > 10$~Mpc---are lower for the higher-redshift filaments (see more discussion about this in Sect. 4.2)

\section{Fits of the radial filament profiles}

Two main methodologies have been used in the literature to quantify the filament width or the boundary of cosmic filaments. The first one is the fitting of analytical functions which provide scales of the galaxy density profiles around filaments \citep[see][]{Tanimura2020, Castignani2022a, Castignani2022b,Bonjean2020, Galarragaespinosa2020}. The second method consists of analyzing the variation of the slope of the galaxy density profile around filaments \citep[see][]{Wang2024}. In this work we use the first approach.

We selected a set of eight different analytical profiles most of them previously used by \citet{Galarragaespinosa2020} to fit the galaxy overdensity around filaments. These profiles effectively capture the general declining trend observed in the data. They include functional forms such as exponential, Gaussian, and polynomial decays, which are flexible and commonly adopted in the literature to describe radial trends in density and mass distributions. The analytical profiles considered are:

\begin{itemize}
    \item Generalized Navarro, Frenk, and White (GNFW) profile 
    \begin{equation}
        \delta_{g,GNFW} (r) = \frac{\delta_{0}}{(\frac{r}{r_{f}})^{\alpha}[1+(\frac{r}{r_{f}})^{\gamma}]^{\frac{\beta-\alpha}{\gamma}}} + \delta_{c},
     \end{equation}
\item Einasto model
\begin{equation}
    \delta_{g,Einasto}(r) = \delta_{0} e^{-(\frac{r}{r_{f}})^{\alpha}}+\delta_{c},
\end{equation}
\item $\beta$-model
\begin{equation}
    \delta_{g,\beta}(r) = \frac{\delta_{0}}{[1+(\frac{r}{r_{f}})^{\alpha}]^{\beta}}+\delta_{c},
\end{equation}
\item Double Power-Law model
\begin{equation}
\delta_{g,PL2} = \frac{\delta_{0}}{[(\frac{r}{r_{f}})^\alpha+(\frac{r}{r_{f}})^{\beta}]}+\delta_{c},
\end{equation}
\item Single Power-Law model
\begin{equation}
\delta_{g,PL1} = \frac{\delta_{0}}{[1+(\frac{r}{r_{f}})^{\alpha}]}+\delta_{c},
\end{equation}
\item Power-Law model
\begin{equation}
\delta_{g,PL3} = \frac{\delta_{0}}{[1+(\frac{r}{r_{f}})^{2}]^{\beta}}+\delta_{c},
\end{equation}

\item Exponential model
\begin{equation}
    \delta_{g,exp}(r) = \delta_{0} e^{-\frac{r}{r_{f}}}+\delta_{c},
\end{equation}
\item Gaussian model
\begin{equation}
    \delta_{g,exp}(r) = \delta_{0} e^{-(\frac{r}{r_{f}})^{2}}+\delta_{c}.
\end{equation}

\end{itemize}

In all of these models, $r_f$ represents the characteristic scale, $\delta_0$ is the density at the filament spine, and $\delta_c$ denotes the constant galaxy overdensity at large distances from the filament spine. The parameters $\alpha$, $\beta$, and $\gamma$ are the exponents that govern the rate of decline in each profile. 

There are relationships among the profiles, as some are special cases of others. For instance, the so-called exponential and Gaussian models are specific cases of the Einasto profile, corresponding to $\alpha = 1$ and $\alpha = 2$, respectively. We will refer to these three models collectively as exponential models, as they all exhibit an exponential decline. The remaining models are referred to as power-law models. This latter group is also internally related—for example, the single power-law profile is a special case of the $\beta$-model. In all these power-law models, the behaviour of the profiles at $r \ll r_{f}$ is nearly constant, while at $r \gg r_{f}$ they exhibit declining trends controlled by the $\alpha$, $\beta$, and $\gamma$ exponents.

We divided the fitted filament profiles into two redshift ranges. The nearest set corresponds to those located within the redshift interval $0.05 < z < 0.1$, while the more distant set includes filaments in the range $0.1 < z < 0.3$. The total number of galaxies located at $D_{fila}<40$ Mpc in the two redshift ranges are $93\,307$ and $131\,330$, respectively. In the following subsections, we present the results of the fitted functions for these two filament sets, along with an analysis of their evolution.

\subsection{Filament profiles at $0.05 < z < 0.1$}

\begin{figure}
    \centering
    \includegraphics[width=0.5\textwidth, trim=0 0 30 0]{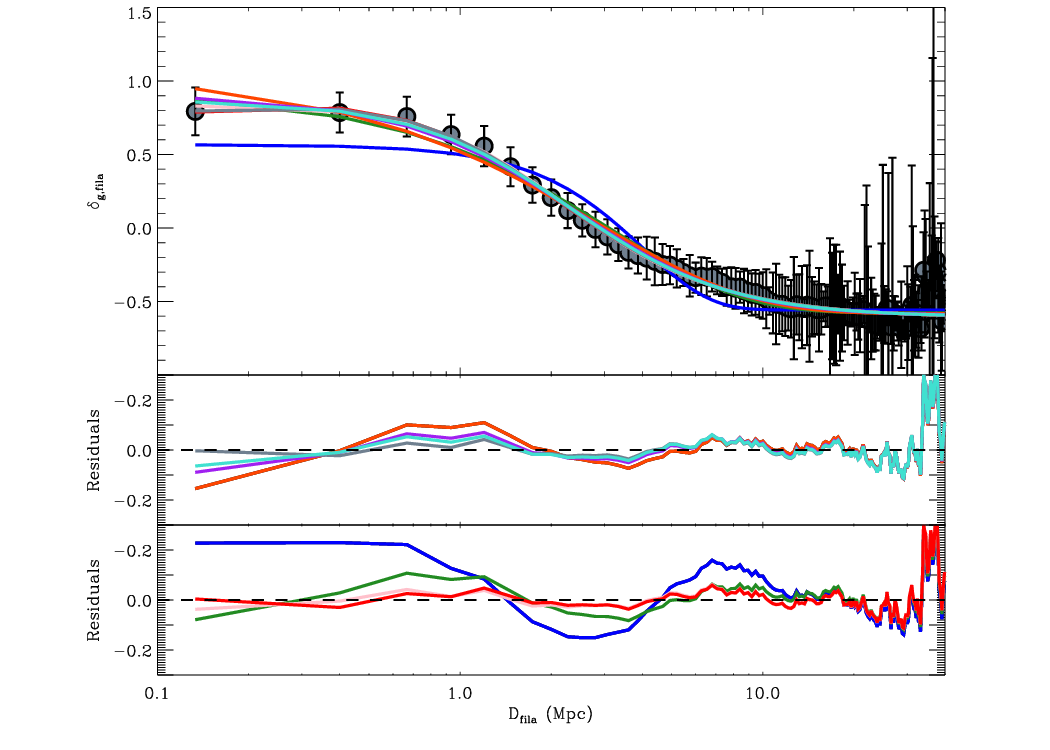}
    \caption{Galaxy overdensity as a function of distance to the filament spine (upper panel). The solid lines represent the best fits using a Gaussian (blue), exponential (green), power-law (pink), GNFW (red), Einasto (orange), double power-law (grey), $\beta-$model (purple), and single power-law (turquoise) models. Middle and lower panels: Residuals between the observed data and the fitted models. Colours correspond to the residuals of the different models as in the upper panel. Uncertainties are similar to those shown in Fig. \ref{fig:overdens_perp}.}
    \label{fig:overdens_perp_fit}
\end{figure}

Figure~\ref{fig:overdens_perp_fit} shows the fits of the different galaxy overdensity profiles to the observed data. The figure also includes the residuals of each fit. The best-fit parameters for each model are listed in Table~\ref{tab:fitted_param_005}. We note that the reduced $\chi^2$ values obtained for the profile fits are significantly below unity. This is not necessarily an indication of overfitting, but rather a consequence of conservative uncertainty estimates. The error bars in the radial profiles were computed as $\sigma/\sqrt{N}$, with $\sigma$ as the standard deviation across a large number of filaments ($N$), which leads to an overestimation of the variance at each radial bin. Therefore, the low reduced $\chi^2$ values reflect this overestimation rather than an excessively complex fit. In this context, comparisons between models based strictly on $\chi^2_\nu$ should be interpreted with caution. We consider the different profiles mainly as flexible functional forms to reproduce the global trend of the galaxy overdensity profiles, rather than to identify a unique physical model. Based on the residuals in Figure~\ref{fig:overdens_perp_fit} and the reduced chi-squared values ($\chi^{2}_{\nu}$) from the fits (see Table~\ref{tab:fitted_param_005}), the exponential models generally yield higher $\chi^{2}_{\nu}$ values. 

The $\beta$-model fit shows a large uncertainty in the characteristic scale, indicating a strong degeneracy in this parameter. The average filament scales derived from the exponential models (excluding the Gaussian model) and the power-law models (excluding the $\beta$-model) are $r_f = 3.06 \pm 0.40$ and $r_f = 2.05 \pm 0.53$ Mpc, respectively. These values suggest that the difference in filament scale between the exponential and power-law models is approximately $1\sigma$. Taking into account both families of models, the average filament scale (excluding the Gaussian and the $\beta$-model) is $r_f = 2.39 \pm 0.69$ Mpc.

Although the constant galaxy overdensity level $\delta_c$ at large distances from the filament spine is expected to be well constrained by the data, we chose to treat it as a free parameter in all fits. This ensures a uniform fitting approach across the various models and redshift bins. As a test, we also performed fits where $\delta_c$ was fixed to the average overdensity at large distances ($D_{\rm fila} > 10$ Mpc). The resulting scale parameters remained broadly consistent. The mean values of the central overdensity and filament scale fixing $\delta_{c}$ turned to be 1.44$\pm$0.27 and 2.97$\pm$0.71. We present the values fitted with constant galaxy overdensity in Tab. \ref{tab:apendix1} at the Appendix A.

\subsection{Filament profiles at $0.1 < z < 0.3$}

\begin{figure}
    \centering
    \includegraphics[width=0.5\textwidth, trim=10 0 30 10]{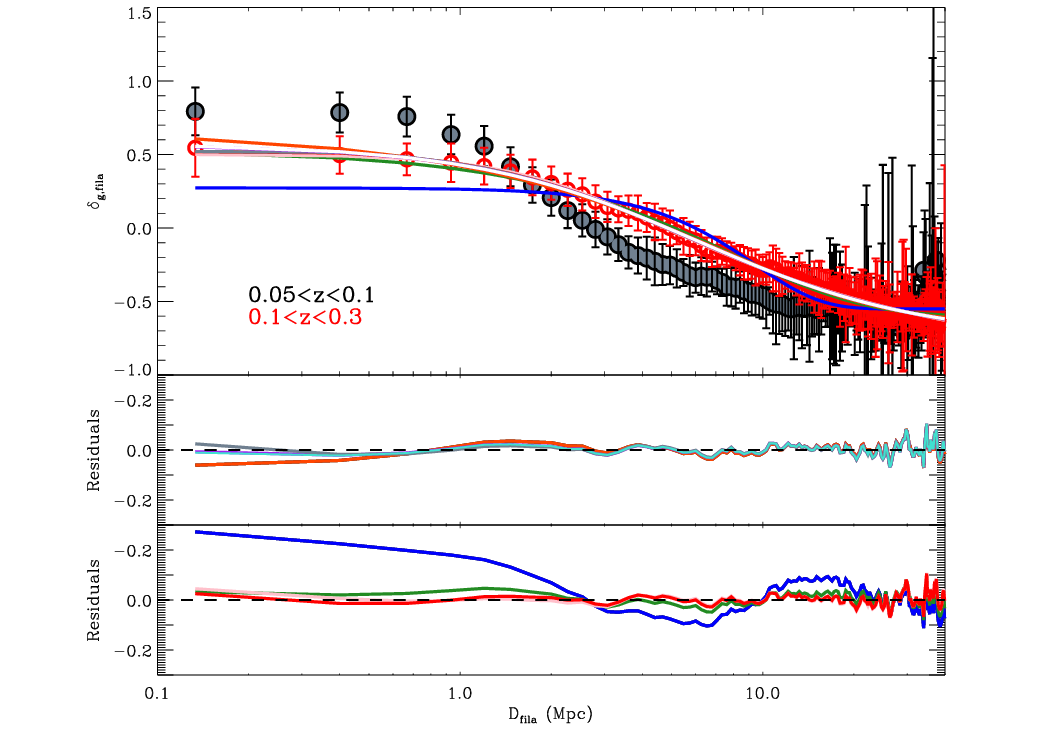}
    \caption{Galaxy overdensity as a function of distance to the filament spine (upper panel). Grey and red points correspond to galaxies in the redshift ranges $0.05<z<0.1$ and $0.1<z<0.3$, respectively. Middle and lower panels: Residuals between the observed data and the fitted models. Uncertainties,  model fits, and residuals colours are similar to those in Fig. \ref{fig:overdens_perp_fit}. }
    \label{fig:overdens_perp_evol}
\end{figure}

Figure~\ref{fig:overdens_perp_evol} shows the galaxy overdensity profiles around filaments located in the redshift range $0.1 < z < 0.3$. For reference, we have also overplotted the galaxy overdensity profile of filaments in the range $0.05 < z < 0.1$. Both profiles exhibit a similar behaviour: a central, approximately constant density up to about 1~Mpc, followed by a decline out to $\sim$10~Mpc. However, filaments at $0.1 < z < 0.3$ display profiles with a lower central galaxy overdensity and a shallower decline compared to those at lower redshift. 

We fitted the same set of profiles used in the previous section. The best-fit parameters for each model are listed in Table~\ref{tab:fitted_param_03}. As in the previous case, the exponential profiles yield higher values of $\chi^{2}_{\nu}$ compared to the power-law fits. 

The exponential models (excluding the Gaussian profile) and the power-law models (excluding the $\beta$ and double power-law profiles) yield mean filament scales of $r_f = 7.78 \pm 0.85$~Mpc and $r_f = 4.10 \pm 1.71$~Mpc, respectively, for the galaxy overdensity profiles of filaments in the redshift range $0.1 < z < 0.3$. Considering both families of models, the average filament scale (excluding the Gaussian, $\beta$ and double power-law models) is $r_f = 5.56 \pm 2.29$~Mpc. In this case, the exponential profiles yield larger filament scales than the power-law models, although the difference is smaller than 2$\sigma$.

\begin{figure}
    \centering
    \includegraphics[width=0.5\textwidth, trim=110 0 140 0]{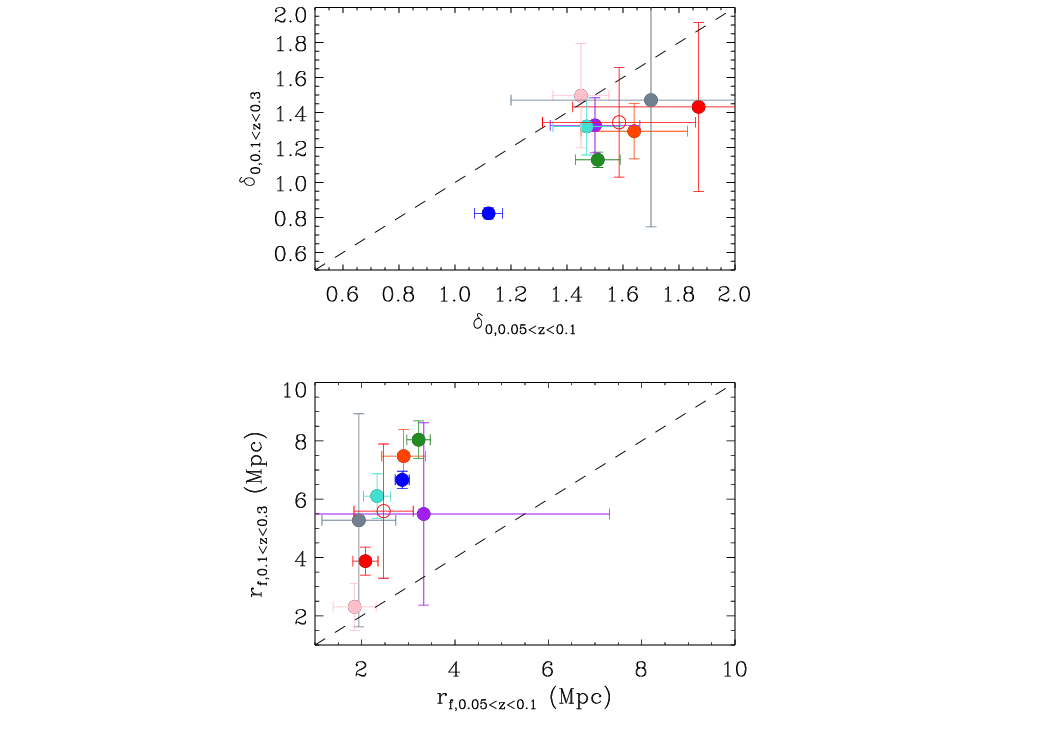}
    \caption{Central galaxy overdensity (upper panel) and filament scales (lower panel) for filaments located at $0.05 < z < 0.1$ and $0.1 < z < 0.3$. Red open symbols indicate the mean values, including uncertainties. The dashed line represents the one-to-one relation of the panels. The colour code of the full points represent the different fitted models as in Fig. \ref{fig:overdens_perp_fit}}
    \label{fig:radius_evol}
\end{figure}

Figure~\ref{fig:radius_evol} shows the central overdensity and the characteristic scales of the filaments in the two redshift ranges considered in this study. Although the differences in central overdensity and scale are smaller than 1$\sigma$ and 2$\sigma$, respectively, this may suggest a mild evolution in the filament parameters between the two redshift intervals.

 In this case, the mean values obtained for $\delta_0$ and $r_f$ when fixing $\delta_c$ are $0.89 \pm 0.18$ and $6.97 \pm 1.34$, respectively (see Tab. \ref{tab:appendix2} in Appendix A). These values are consistent, within the uncertainties, with those obtained when $\delta_c$ was treated as a free parameter.

\subsection{Evolution of the filaments with redshift}
\label{evolfil}

It is important to note that the observed evolution of filament properties with redshift may be affected by several observational biases \citep[see e.g.][]{Malavasi2017, Laigle2018, Kraljic2018}. A potential bias that could account for the observed evolution is the use of physical rather than comoving distances in the derivation of the filament overdensity profiles. These two distance measures are related by the expression $D_{\mathrm{phys}} = D_{\mathrm{com}} / (1 + z)$, where $D_{phys}$ and $D_{com}$ are the physical and comovil distances, respectively. Physical distances provide a view of the filaments at a given cosmic time, but do not correct for the expansion of the Universe, which is essential when comparing structures across redshift. However, the two redshift bins analysed in this work are relatively close. The mean redshift of the galaxies is $z = 0.07$ for the $0.05 < z < 0.1$ interval, and $z = 0.14$ for the $0.1 < z < 0.3$ interval. The comoving filament widths measured are $D_{\mathrm{fila,com}} = 2.56$~Mpc and $6.34$~Mpc for the low- and high-redshift bins, respectively. If the evolution of the filament width were driven solely by cosmic expansion, the expected scaling factor between the comoving widths would be $(1 + 0.07)/(1 + 0.14) \approx 0.94$. Instead, we observe a ratio of $2.55/6.35 \approx 0.4$, which is significantly smaller than the expected value. Moreover, if cosmic expansion alone were responsible for the evolution, the physical width of the filaments should decrease with redshift (i.e., filaments at lower redshift should appear physically larger than those at higher redshift). In our case, we observe the opposite trend. This discrepancy strongly suggests that the evolution in filament width cannot be explained by cosmic expansion alone, and instead reflects an intrinsic physical evolution of the filamentary structures. 

Observations always exhibit a redshift dependence in the sampling of galaxies, and the main spectroscopic sample of the SDSS survey used in this work is not exempt from this effect. In particular, its completeness is a function of redshift, with higher completeness for nearby galaxies than for more distant ones. This effect leads to observed galaxy densities being higher at lower redshifts than at higher ones. In contrast, distances to filaments would appear larger at higher redshifts due to the redshift completeness effect \citep[see Fig. 7, ][]{Chen2016}. These effects can lead to a systematic underestimation of filament contrast and an over-smoothing of their density profiles.

This observational bias in galaxy overdensity and distances to filaments can be addressed by dividing these two quantities by their mean values at the redshift of the galaxies. The galaxy overdensities used in this work are taken from \citet{Zarattini2025}, in which such a correction is already taken into account (see the definition that we reported in Sect. \ref{sec:filaments}). However, this correction has not been applied to the filament scales reported previously.

To determine whether this effect needs to be considered in our analysis, we have computed the mean distances of galaxies to filaments as a function of redshift. This quantity strongly depends on redshift, varying from 4 Mpc at $z=0.05$ to 20 Mpc at $z=0.3$. Similar values were obtained by \citet{Chen2016}. To address this effect, we have obtained the galaxy overdensity profiles by scaling $D_{fila}$ to its mean value at the redshift of the galaxies. This mitigates the impact of redshift-dependent incompleteness on the galaxy-filament distance distributions and has been used in the literature \citep[see e.g.][]{Donnan2022}. By applying this scaling, we reduce biases when comparing overdensity profiles across different redshift slices. However, we acknowledge that this approach does not fully correct for all effects of incompleteness, particularly the potential missing filaments at higher redshift.

\begin{figure}
    \centering
    \includegraphics[width=0.5\textwidth, trim=10 0 30 10]{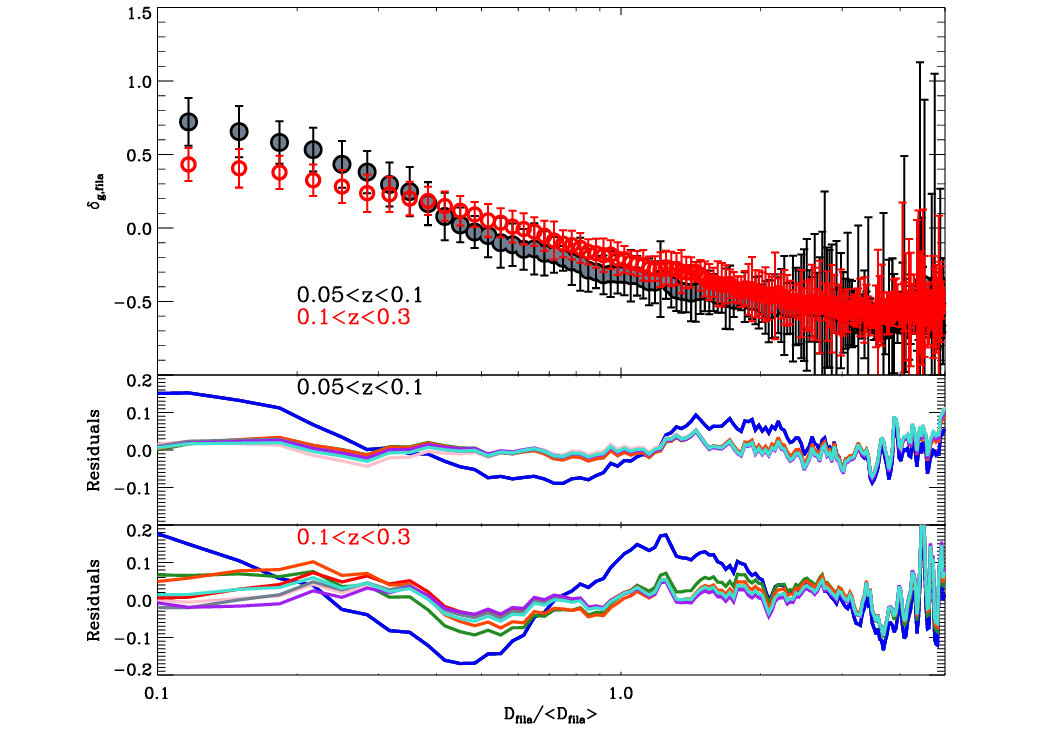}
    \caption{Galaxy overdensity as a function of the normalised distance to the filament skeleton (top panel). Grey and red points represent the overdensity profiles for galaxies in the redshift ranges $0.05 < z < 0.1$ and $0.1 < z < 0.3$, respectively. The residuals of the fits are displayed in the middle panel for filaments at $0.05 < z < 0.1$, and in the bottom panel for filaments at $0.1 < z < 0.3$. The colour coding of both residuals and uncertainties follows the same scheme as in Fig.~\ref{fig:overdens_perp_fit}.}
    \label{fig:overdens_perp_evol_scaled}
\end{figure}

Figure \ref{fig:overdens_perp_evol_scaled} shows the galaxy overdensity profiles of filaments in the two redshift ranges, with distances normalised by the mean galaxy–filament distance at each redshift. Similar to the profiles shown in Fig. \ref{fig:overdens_perp_evol}, filaments at $0.1 < z < 0.3$ exhibit lower central overdensities and broader scales compared to those at $0.05 < z < 0.1$. Tables \ref{tab:fitted_param_scaled_005} and \ref{tab:fitted_param_scaled_03} list the fitted parameters for the different functional forms used. 

For the scaled density profiles of filaments at $0.05 < z < 0.1$, the Gaussian model provides the poorest fit. Two models—the GNFW and the Double Power-law—yield filament scales with large uncertainties.  The mean filament scale derived from the exponential models (excluding the Gaussian) and the power-law models (excluding the GNFW and Double Power-law) are $0.54 \pm 0.08$ and $0.29 \pm 0.10$, respectively. Considering all models (excluding the Gaussian, GNFW and the Double Power-law), the mean filament scale at $0.05 < z < 0.1$ is $0.39 \pm 0.15$.

For the overdensity profile of filaments at $0.1 < z < 0.3$, the Gaussian model again provides the poorest fit to the observations. However, in this case, the exponential models offer better fits than the power-law ones, with the best fit given by the exponential model. The mean filament scale values obtained from the exponential models (excluding the Gaussian), power-law models, and all well-constrained models are $0.91 \pm 0.08$, $0.85 \pm 0.38$, and $0.87 \pm 0.33$, respectively.

\begin{figure}
    \centering
    \includegraphics[width=0.5\textwidth, trim=120 0 140 0]{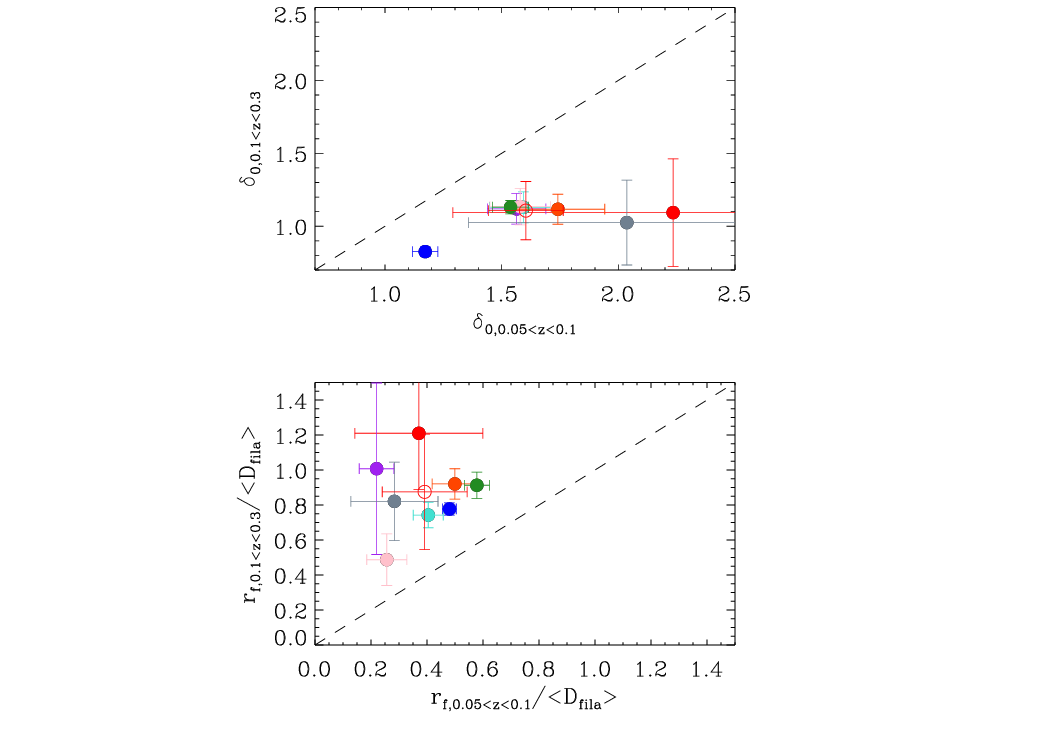}
    \caption{Central galaxy overdensity (upper panel) and normalised filament scales (lower panel) for filaments located at $0.05 < z < 0.1$ and $0.1 < z < 0.3$. Red open symbols indicate the mean values, including uncertainties. The dashed line represents the one-to-one relation of the panels. The color code of the full points represent the different fitted models as in Fig. \ref{fig:overdens_perp_fit}}
    \label{fig:radius_norm_evol}
\end{figure}

Figure \ref{fig:radius_norm_evol} shows the comparison of the galaxy overdensity at the filament spine and the filament scale for the two redshift ranges considered. Although the differences are smaller than 2$\sigma$, filaments at $0.1 < z < 0.3$ exhibit lower central galaxy overdensities and larger scales compared to those at $0.05 < z < 0.1$. This result suggests a mild evolution in the properties of filaments with redshift, indicating that the evolution observed in the scales of the unscaled filament profiles cannot be explained by the redshift completeness of the observational data used in this study.

The relatively narrow redshift range explored in this study implies that the differences observed in central galaxy overdensity and filament scale between the two redshift bins remain modest. To robustly confirm the redshift evolution of filament properties, a more extensive analysis covering a broader redshift range would be required.

\begin{table*}[]
    \caption{Best-fit parameters of the observed galaxy overdensity profiles around filaments at $0.05 < z < 0.1$.}
        \label{tab:fitted_param_005}
    \centering
    \begin{tabular}{cccccccc}
    \hline
     Profile & $\delta_{0}$ & $r_{f}$ (Mpc) & $\alpha$ & $\beta$ & $\gamma$ & $\delta_{c}$ & $\chi^{2}_{\nu}$ \\
     \hline
     Exponential profiles \\
     \hline
     Einasto & 1.64$\pm$0.19 & 2.90$\pm$0.47 & 0.87$\pm$0.13 &  &  & -0.58$\pm$0.01 & 0.368\\
     Exponential & 1.51$\pm$0.08 & 3.22$\pm$0.25 & 1.00 &  & & -0.58$\pm$0.01 & 0.374\\
     Gaussian & 1.12$\pm$0.05 & 2.87$\pm$0.15 & 2.00 &  &  & -0.56$\pm$0.01 & 0.618\\
     \hline
     Power-law profiles \\
     \hline
     GNFW & 1.87$\pm$0.45 & 2.08$\pm$0.27 & -0.10$\pm$0.10 & 3.81$\pm$0.42 & 1.42$\pm$0.34 & -0.61$\pm$0.02 & 0.366\\
     $\beta$ model & 1.50$\pm$0.16 & 3.33$\pm$3.98 & 1.39$\pm$0.66 & 1.52$\pm$1.93 &  & -0.59$\pm$0.03 & 0.366\\
     Single Power-Law & 1.47$\pm$0.12 & 2.33$\pm$0.29 & 1.66$\pm$0.22 & 1.00 &  & -0.60$\pm$0.03  & 0.364\\
     Power-Law & 1.45$\pm$0.10 & 1.85$\pm$0.46 & 2.00 & 0.70$\pm$0.175 &  & -0.61$\pm$0.03 & 0.365\\
     Double Power-Law & 1.70$\pm$0.50 & 1.94$\pm$0.79 & -0.07$\pm$0.14 & 1.53$\pm$0.34 & & -0.61$\pm$0.02 & 0.365\\
     \hline
     \end{tabular}
     \tablefoot{Values without uncertainties correspond to specific cases where one or more parameters were fixed during the fitting procedure.}
\end{table*}

\begin{table*}[]
    \caption{Best-fit parameters of the observed galaxy overdensity profiles around filaments at $0.1 < z < 0.3$.}
        \label{tab:fitted_param_03}
    \centering
    \begin{tabular}{cccccccc}
    \hline
     Profile & $\delta_{0}$ & $r_{f}$ (Mpc) & $\alpha$ & $\beta$ & $\gamma$ & $\delta_{c}$ & $\chi^{2}_{\nu}$ \\
     \hline
     Exponential profiles \\
    \hline
     Einasto & 1.29$\pm$0.16 & 7.47$\pm$0.92 & 0.80$\pm$0.14 &  &  & -0.64$\pm$0.04 & 0.0452\\
     Exponential & 1.13$\pm$0.04 & 8.04$\pm$0.65 & 1.00 &  & & -0.60$\pm$0.02 & 0.0588\\
     Gaussian & 0.82$\pm$0.03 & 6.66$\pm$0.29 & 2.00&  &  & -0.55$\pm$0.01 & 0.3389\\
     \hline
     Power-law profiles \\
     \hline
     GNFW & 1.43$\pm$0.48 & 3.87$\pm$0.48 & 0.84$\pm$0.21 & 3.01$\pm$0.42 & -1.27$\pm$0.58 & -0.81$\pm$0.09 & 0.0419\\
     $\beta$ model & 1.33$\pm$0.16 & 5.49$\pm$3.13 & 1.21$\pm$0.30 & 0.90$\pm$0.52 &  & -0.76$\pm$0.10 & 0.0419\\
     Single Power-Law & 1.32$\pm$0.16 & 6.11$\pm$0.77 & 1.17$\pm$0.22 & 1.00 &  & -0.75$\pm$0.08 & 0.0417\\
     Power-Law & 1.50$\pm$0.30 & 2.31$\pm$0.81 & 2.00 & 0.25$\pm$0.11 &  & -0.99$\pm$0.26 & 0.0426\\
     Double Power-Law & 1.47$\pm$0.72 & 5.28$\pm$3.65 & 1.08$\pm$0.46 & -0.03$\pm$0.14 & & -0.77$\pm$0.13 & 0.0416\\
     \hline
     \end{tabular}
     \tablefoot{Values without uncertainties correspond to specific cases where one or more parameters were fixed during the fitting procedure.}
\end{table*}

\begin{table*}[]
    \caption{Best-fit parameters of the normalised galaxy overdensity profiles around filaments at $0.05 < z < 0.1$.}
        \label{tab:fitted_param_scaled_005}
    \centering
    \begin{tabular}{cccccccc}
    \hline
     Profile & $\delta_{0}$ & $r_{f}$ & $\alpha$ & $\beta$& $\gamma$ & $\delta_{c}$ & $\chi^{2}_{\nu}$ \\
     \hline
     Exponential profiles \\
     \hline
     Einasto & 1.74$\pm$0.20 & 0.50$\pm$0.08 & 0.81$\pm$0.13 &  &  & -0.62$\pm$0.02 & 0.1392\\
     Exponential & 1.54$\pm$0.08 & 0.58$\pm$0.04 & 1.00 &  & & -0.60$\pm$0.02 & 0.1551\\
     Gaussian & 1.17$\pm$0.05 & 0.48$\pm$0.02 & 2.00&  &  & -0.57$\pm$0.01 & 0.4072\\
     \hline
     Power-law profiles \\
     \hline
     GNFW & 2.23$\pm$0.94 & 0.37$\pm$0.23 & 1.51$\pm$0.54 & -1.26$\pm$0.41 & -1.14$\pm$0.51 & -0.67$\pm$0.04 & 0.1193\\
     $\beta$-model & 1.56$\pm$0.13 & 0.22$\pm$0.06 & 2.58$\pm$1.04 & 0.35$\pm$0.22 &  & -0.73$\pm$0.07 & 0.1144\\
     Single Power-Law & 1.59$\pm$0.15 & 0.40$\pm$0.05 &  1.46$\pm$0.22 & 1.00 & & -0.66$\pm$0.03 & 0.1200\\
     Power-Law & 1.58$\pm$0.13 & 0.26$\pm$0.07 & 2.00 & 0.51$\pm$0.14 &  & -0.71$\pm$0.06 & 0.1149\\
     Double Power-Law & 2.04$\pm$0.68 & 0.28$\pm$0.16 & 1.25$\pm$0.33 & -0.11$\pm$0.15 & & -0.68$\pm$0.05 & 0.1161\\
     \hline
     \end{tabular}
     \tablefoot{Values without uncertainties correspond to specific cases where one or more parameters were fixed during the fitting procedure.}
\end{table*}

\begin{table*}[]
    \caption{Best-fit parameters of the normalised galaxy overdensity profiles around filaments at $0.1 < z < 0.3$.}
        \label{tab:fitted_param_scaled_03}
    \centering
    \begin{tabular}{cccccccc}
    \hline
     Profile & $\delta_{0}$ & $r_{f}$ & $\alpha$ & $\beta$ & $\gamma$ & $\delta_{c}$ & $\chi^{2}_{\nu}$ \\
     \hline
     Exponential profiles \\
     \hline
     Einasto & 1.12$\pm$0.10 & 0.92$\pm$0.09 & 1.02$\pm$0.15 &  &  & -0.57$\pm$0.03 & 0.0707\\
     Exponential & 1.13$\pm$0.05 & 0.91$\pm$0.07 & 1.00 &  & & -0.58$\pm$0.02 & 0.0704\\
     Gaussian & 0.83$\pm$0.03 & 0.78$\pm$0.04 & 2.00&  &  & -0.54$\pm$0.01 & 0.2525\\
     \hline
     Power-law profiles \\
     \hline
     GNFW & 1.09$\pm$0.37 & 1.21$\pm$0.32 & 2.29$\pm$0.40 & -1.08$\pm$0.27 & -1.34$\pm$0.41 & -0.62$\pm$0.03 & 0.0788\\
     $\beta$-model & 1.12$\pm$0.10 & 1.01$\pm$0.49 & 1.36$\pm$0.38 & 1.44$\pm$0.80 &  & -0.62$\pm$0.06 & 0.0812\\
     Single Power-Law & 1.13$\pm$0.11 & 0.74$\pm$0.07 &  1.52$\pm$0.25 & 1.00 & & -0.65$\pm$0.05 & 0.0815\\
     Power-Law & 1.13$\pm$0.13 & 0.49$\pm$0.15 & 2.00 & 0.53$\pm$0.21 &  & -0.69$\pm$0.09 & 0.0887\\
     Double Power-Law & 1.03$\pm$0.29 & 0.82$\pm$0.22 & 1.65$\pm$0.46 & 0.03$\pm$0.09 & & -0.64$\pm$0.06 & 0.0813\\
     \hline
     \end{tabular}
     \tablefoot{Values without uncertainties correspond to specific cases where one or more parameters were fixed during the fitting procedure.}
\end{table*}

\section{Discussion}

Our results confirm that cosmic filaments exhibit galaxy overdensity profiles that decline with increasing perpendicular distance from the filament spine, consistent with previous studies based on both observations and cosmological simulations \citep[e.g.][]{Tanimura2020, Galarragaespinosa2020, Bonjean2020}. Moreover, a mild evolution with redshift is found. Both aspects are discussed in the rest of this Section. 

\subsection{Comparison of filament scales}

We find that the characteristic scale of the galaxy overdensity profile is approximately $2.39 \pm 0.69$ Mpc for filaments in the redshift range $0.05 < z < 0.1$, and $5.56 \pm 2.29$ Mpc for filaments at $0.1 < z < 0.3$. These measurements are broadly consistent with prior observational estimates. For example, \citet{Castignani2022b} studied 13 filaments near the Virgo cluster and reported characteristic scales from 0.25 to 3.79 Mpc (converted to our cosmology), while \citet{Castignani2022a} found a mean scale of $3.27 \pm 0.79$ Mpc when considering all filaments together. Both results align with our measurements in the lower-redshift bin. In contrast, \citet{Wang2024} found a smaller scale of $\sim$1 Mpc at $z=0$, attributing the discrepancy to their redshift-space distortion (RSD) correction. Without this correction, their estimate increased to $\sim$2 Mpc, closer to our low-redshift measurement.

At higher redshifts ($0.1 < z < 0.3$), \citet{Bonjean2020} reported a characteristic scale of $7.3 \pm 0.2$ Mpc (converted to our cosmology), in agreement with our findings and significantly larger than the Virgo cluster filaments. 

Simulations of cosmic filaments in DM only cosmologies typically yield narrower widths, around 0.1–0.6 Mpc \citep{Colberg2005, Lemson2006, Navarro1996, Springel2018, Henning2017}. Hydrodynamical simulations, which include baryonic physics, predict broader filaments, with typical widths of 1–3 Mpc \citep[e.g.][]{Pillepich2018, Schaye2015, Dave2019, Dubois2014, Puchwein2013, Kuchner2020}. Using analytical profile fitting, \citet{Galarragaespinosa2020} found filament scales of 0.14–0.25 Mpc across several simulations (Illustris-TNG, Illustris, Magneticum), regardless of filament length.

The consistency between our low-redshift measurements and simulations that include baryonic processes underscores the influence of thermal pressure, stellar and AGN feedback, and gas accretion in shaping filament widths. However, the relatively small widths reported by \citet{Galarragaespinosa2020}—compared to those from both observations and other simulations—could reflect either observational biases or limitations in capturing the full physics of large-scale structure formation.

\subsection{Evolution of filament scales with redshift}

We observe a mild but significant evolution in filament properties between the two redshift bins. Lower-redshift filaments appear more centrally concentrated and exhibit smaller characteristic scales than their higher-redshift counterparts. This trend may account for differences reported in the literature, such as the smaller filament widths near the Virgo cluster \citep{Castignani2022a, Castignani2022b} relative to the broader filaments reported at $0.1 < z < 0.3$ \citep{Bonjean2020}. These comparisons highlight the importance of redshift coverage when interpreting filament measurements.

The observational variation of the width of the filament with redshift at $z<1$  was already observed. \citet{Choi2010} found that, while filament lengths remained roughly constant from $z = 0.8$ to $z = 0.1$, their widths decreased with cosmic time. This indicates that since $z\sim 1$ filaments have not collapsed in their longitudial direction but experience collapse in their perpendicular dimension. This is  consistent with anisotropic gravitational collapse \citep{Bond2010}.

\begin{figure}
    \centering
    \includegraphics[width=0.5\textwidth] {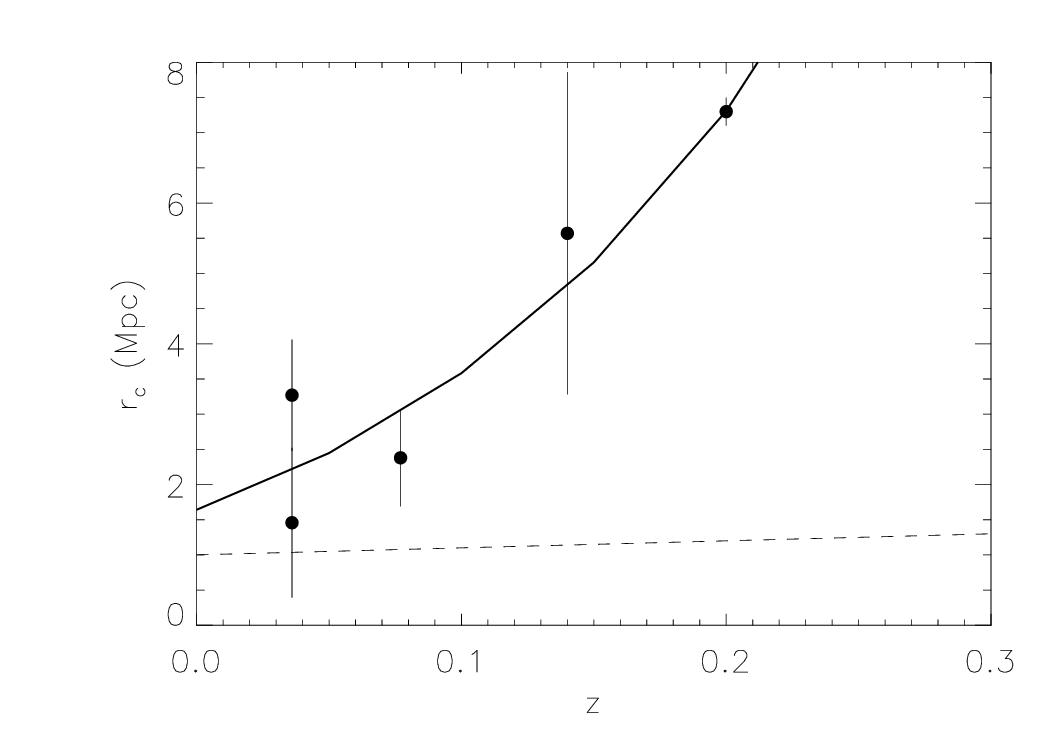}
    \caption{Scales of the filament densities from literature as a function of redshift. The points represents the filament scales from: Virgo cluster ($z=0.036$), this work ($z=0.07$ and $z=0.14$), and \cite{Bonjean2020} ($z=0.2$) The solid line represents the fit $r_{fila} \propto (1+z)^{\alpha}$, the dashed-line shows a linear relation with $(1+z)$.}
    \label{fig:rfila_z}
\end{figure}

Figure~\ref{fig:rfila_z} summarizes filament scale measurements from this work and previous studies, plotted as a function of redshift. The values include those from \citet{Castignani2022a, Castignani2022b}, \citet{Bonjean2020}, and our two redshift bins,for which we use the mean redshift of each interval. The data reveal an evolution that exceeds a simple linear dependence on $(1+z)$; a power-law fit yields $r_{\text{fila}} \propto (1+z)^{8.19 \pm 1.64}$, much steeper than predictions from simulations \citep[e.g.][]{Cautun2014, Gheller2016}.

The observed increase in filament width—from $2.39 \pm 0.69$ Mpc at $z \sim 0.075$ to $5.56 \pm 2.29$ Mpc at $z \sim 0.2$—could place strong constraints on the interplay between gravitational collapse and cosmic expansion. Under the $\Lambda$CDM model, the growth of cosmic structures slows down at late times as dark energy becomes the dominant component of the Universe.  A narrowing of filaments at low $z$ aligns with this scenario, reflecting continued local collapse despite cosmic acceleration. Conversely, broader filaments at higher $z$ may reflect less evolved, diffuse structures. Such trends can complement traditional large-scale structure probes such as redshift-space distortions or halo mass functions \citep[e.g.][]{Peebles1980, Percival2005}, and discrepancies could hint at new physics, such as evolving dark energy or modified gravity \citep{Llinares2014, Leonard2015}.

We note that while our observed trend may seem to contradict simulation-based studies, such as \citet{Ramsoy2021}, who find that filament core radii decrease with redshift as $r_{\text{core}} \propto (1+z)^{-3.18}$, this discrepancy arises from structural definitions. Their measurements refer to compact DM or gas cores, while our analysis traces the broader distribution of galaxies around filaments. At high redshift, galaxies are more diffusely distributed, whereas at low redshift gravitational collapse leads to more concentrated, denser structures. This contrast underscores the need to distinguish between the dense filament core and its surrounding galaxy envelope when comparing simulations and observations.

Numerical simulations do not show this evolution at $z<1$ in the width of the filaments. \citet{Wang2024} reported that most of the evolution in filament width occurred between $z = 4$ and $z = 1$, with minimal change thereafter.  The low-redshift evolution could inform how filamentary environments regulate galaxy growth and star formation \citep{Kraljic2018, Laigle2018}. Confirming the observed evolutionary trend in the filament properties will require a broader observational baseline. A significant evolution in filament properties at $z < 0.3$ would imply ongoing structure formation in the late Universe, despite accelerated expansion \citep{Bond1996, Cautun2014}. However, the discrepancy between our observing results and numerical simulations may point out that current state-of-the-art simulations may not fully capture the relevant physical processes at low redshifts. Simulations may lack critical baryonic processes—such as gas heating, stellar and AGN feedback, and delayed gas accretion—that could broaden filaments in the real world. In this context, the evolution we detect may reflect a genuine physical effect currently underrepresented in simulations \citep[e.g.][]{Dubois2014, Pillepich2018, Martizzi2019}.

\subsection{Possible observational bias}

Possible explanations for the observed broadening include observational biases such as redshift uncertainties, projection effects, and reduced spatial resolution at higher $z$, which could artificially smooth filament profiles. We mitigated this effect (see Sect.~\ref{evolfil}) by normalizing distances by the mean galaxy–filament separation at each redshift, yet the trend persists in the normalised profiles.  However, with only two redshift bins, the data remain insufficient to robustly model this evolution. Additional measurements are needed to draw firm conclusions.

Another potential observational bias arises from the spectroscopic completeness limit of the main spectroscopic sample of the SDSS-DR16 survey, which only includes galaxies down to an apparent magnitude of $r_p = 17.77$. As a consequence, the galaxy overdensity profiles of filaments in the two redshift intervals considered in this work are traced by different galaxy populations in terms of stellar mass. In particular, filaments in the range $0.1 < z < 0.3$ are primarily traced by more massive galaxies compared to those in the $0.05 < z < 0.1$ interval. To assess the impact of this mass-dependent selection on the observed evolution of filament scales and central overdensity, we selected galaxies with $\log(M/M_\odot) > 10.8$ within the redshift ranges $0.05 < z < 0.1$ and $0.1 < z < 0.2$. These redshift and mass cuts ensure a mass-complete sample in both intervals, allowing for a fair comparison of the filament properties traced by galaxies of similar mass.

\begin{figure}
    \centering
    \includegraphics[width=0.5\textwidth] {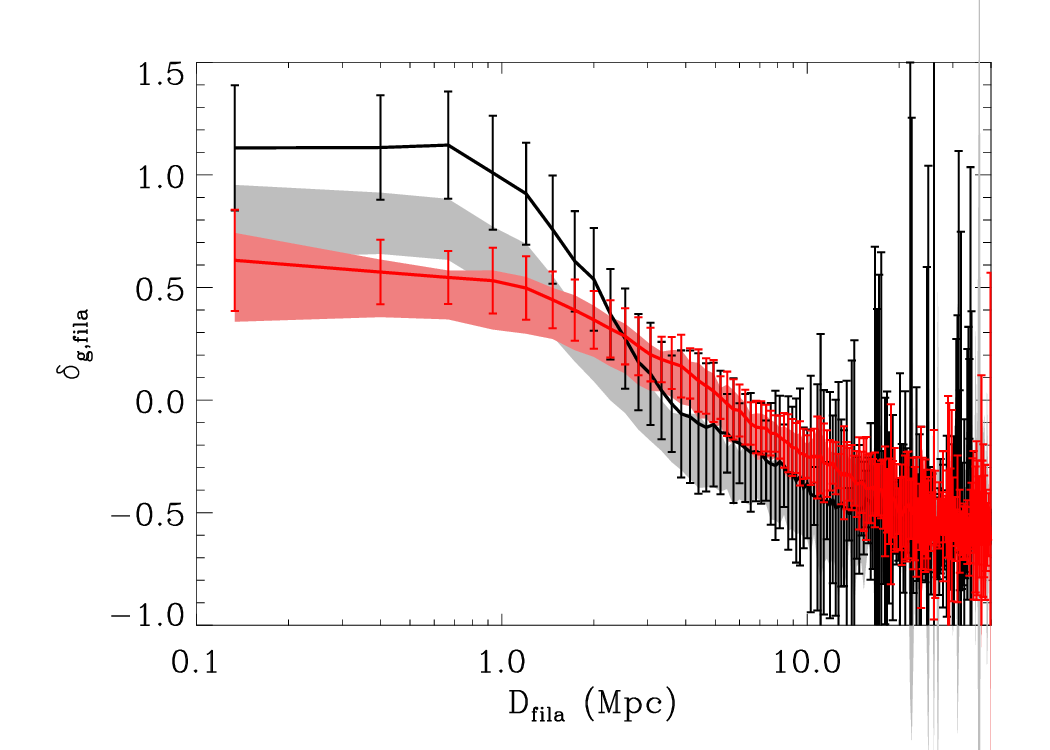}
    \caption{Galaxy overdensity as a function of the distance to the filament spine. The grey and pink shaded regions correspond to the original profiles shown in Fig.~\ref{fig:overdens_perp_evol}, computed without applying any stellar mass cut. The black and red solid lines represent the galaxy overdensity profiles for stellar-mass-selected galaxies with $\log(M/M_\odot) > 10.8$ in the redshift intervals $0.05 < z < 0.1$ and $0.1 < z < 0.2$, respectively.}
    \label{fig:evol_masscut}
\end{figure}

Figure~\ref{fig:evol_masscut} presents the comparison between the galaxy overdensity profiles of filaments with and without a stellar mass cut in the two redshift intervals. For filaments in the range $0.05 < z < 0.1$, the shape of the profiles obtained with both the full and mass-selected galaxy samples are remarkably similar. The main difference lies in the amplitude: more massive galaxies tend to reside in regions of higher overdensity, with deviations at the level of $\sim1\sigma$. This mass segregation of galaxies in filaments has been reported in other observational studies \citep[see, e.g.][]{Chen2017}. For the redshift interval $0.1 < z < 0.2$, the overdensity profile of galaxies with $\log(M/M_\odot) > 10.8$ is consistent, within uncertainties, with that of the complete sample. Importantly, the redshift evolution observed in the width and central overdensity of filament profiles persists even when applying the mass cut. This indicates that the observed evolution could be not driven by a Malmquist bias associated with the spectroscopic completeness limit of the SDSS data.

We note that although our galaxy samples have been selected to be mass-complete within each redshift bin, the filament catalogue used in this study is based on the original spectroscopic sample from \citet{Chen2016}, which suffers from incompleteness at higher redshifts. This inherent limitation means that the filament network itself may be biased due to incomplete filament identification, particularly at higher redshifts. Ideally, a re-identification of filaments using mass-complete samples would be necessary to fully mitigate this bias; however, this is beyond the scope of the present work given our reliance on a public filament catalogue. We therefore caution that some systematic differences in filament detection completeness between redshift intervals could affect the comparison of filament properties across redshift. Future studies with improved filament catalogues are needed to fully address this issue.

\section{Conclusions}

In this paper, we have analysed the galaxy overdensity profiles around filaments located in two redshift intervals: $0.05 < z < 0.1$ and $0.1 < z < 0.3$. The main conclusions of this study are summarized as follows:

\begin{itemize}
    \item The radial galaxy overdensity profiles, measured in the direction perpendicular to the filament spine, remain approximately constant within $D_{\mathrm{fila}} < 1$~Mpc. Beyond this distance, the density gradually declines and reaches an approximately constant background level at $D_{\mathrm{fila}} \sim 10$~Mpc.

    \item The galaxy overdensity along the filament axis remains nearly constant for distances larger than 20~Mpc from the nearest intersection. Furthermore, a clear overdensity gradient is observed as a function of the perpendicular distance to the filament. In particular, galaxies located near the filament spine ($D_{\mathrm{fila}} < 1$~Mpc) experience a local density that is approximately 2.5 times higher than that of galaxies at $D_{\mathrm{fila}} \sim 5$~Mpc.

    \item Different mathematical functions have been fitted to the galaxy overdensity profiles around filaments at two redshift ranges: $0.05 < z < 0.1$ and $0.1 < z < 0.3$. The mean physical widths of the filaments derived from these fits are $r_{f} = 2.39 \pm 0.69$ Mpc and $r_{f} = 5.56 \pm 2.29$ Mpc, respectively.  These results suggest a mild evolution in the characteristic width of filaments across the two redshift intervals.

    \item We have constructed normalised galaxy overdensity profiles by scaling the galaxy–filament distances with the mean galaxy distance to filaments at a given redshift. This approach minimises redshift-dependent observational biases due to incompleteness. The normalised filament widths obtained for $0.05 < z < 0.1$ and $0.1 < z < 0.3$ are $r_{f} = 0.39 \pm 0.15$ and $r_{f} = 0.87 \pm 0.33$, respectively. These values confirm the presence of a mild redshift evolution in the filament density profiles even after correcting for redshift-related selection effects.
\end{itemize}

Future work will extend this analysis by exploring the dependence of filament density profiles in a wider redshift range. In addition, galaxy properties will be analysed as a function of redshift in the influence zone of the filaments. This will require deeper redshift surveys (e.g. BOSS, eBOSS, DESI) to better trace the evolution of filament scales up to higher redshifts.

\begin{acknowledgements}

 We would like to thank the anonymous referee for the careful reading of the manuscript and for the valuable and constructive comments provided. SZ and JALA acknowledge financial support provided by the Spanish Ministerio de Ciencia, Innovación y Universidades (MICIU) through the project PID2023-153342NB-I00. SZ acknowledges the financial support provided by the Governments of Spain and Arag\'on through their general budgets and the Fondo de Inversiones de Teruel, the Aragonese Government through the Research Group E16\_23R, and the Spanish Ministry of Science and Innovation and the European Union - NextGenerationEU through the Recovery and Resilience Facility project ICTS-MRR-2021-03-CEFCA. Funding for the Sloan Digital Sky Survey IV has been provided by the Alfred P. Sloan Foundation, the U.S. Department of Energy Office of Science, and the Participating Institutions. SDSS acknowledges support and resources from the Center for High-Performance Computing at the University of Utah. The SDSS web site is www.sdss4.org. SDSS is managed by the Astrophysical Research Consortium for the Participating Institutions of the SDSS Collaboration including the Brazilian Participation Group, the Carnegie Institution for Science, Carnegie Mellon University, Center for Astrophysics Harvard \& Smithsonian (CfA), the Chilean Participation Group, the French Participation Group, Instituto de Astrofísica de Canarias, The Johns Hopkins University, Kavli Institute for the Physics and Mathematics of the Universe (IPMU) / University of Tokyo, the Korean Participation Group, Lawrence Berkeley National Laboratory, Leibniz Institut für Astrophysik Potsdam (AIP), Max-Planck-Institut für Astronomie (MPIA Heidelberg), Max-Planck-Institut für Astrophysik (MPA Garching), Max-Planck-Institut für Extraterrestrische Physik (MPE), National Astronomical Observatories of China, New Mexico State University, New York University, University of Notre Dame, Observatório Nacional / MCTI, The Ohio State University, Pennsylvania State University, Shanghai Astronomical Observatory, United Kingdom Participation Group, Universidad Nacional Autónoma de México, University of Arizona, University of Colorado Boulder, University of Oxford, University of Portsmouth, University of Utah, University of Virginia, University of Washington, University of Wisconsin, Vanderbilt University, and Yale University. 
\end{acknowledgements}

\bibliography{bibliografia}

\begin{appendix}
\onecolumn
\section{Galaxy density profile fits with fixed large-radius overdensity}

In this appendix we present the full set of galaxy density–profile fits obtained when fixing the large-radius galaxy overdensity, $\delta_{c}$, to the mean background value. These fits were performed in parallel to the unconstrained analysis used throughout the main text and are included here for comparison. The fit for some of the profiles with fixed background did not converge or produce none physical fits, we leave these fits empty in the tables.

\begin{table*}[h]
    \caption{Best-fit parameters of the observed galaxy overdensity profiles around filaments at $0.05 < z < 0.1$. }
        \label{tab:apendix1}
    \centering
    \begin{tabular}{ccccccc}
    \hline
     Profile & $\delta_{0}$ & $r_{f}$ (Mpc) & $\alpha$ & $\beta$ & $\gamma$ &  $\chi^{2}_{\nu}$ \\
     \hline
     Exponential profiles \\
     \hline
     Einasto & 1.57$\pm$0.16 & 2.86$\pm$0.42 & 0.95$\pm$0.13 &  &  &  0.403\\
     Exponential & 1.52$\pm$0.08 & 2.99$\pm$0.08 & 1.00 &  &  & 0.428\\
     Gaussian & 1.13$\pm$0.05 & 2.80$\pm$0.15 & 2.00 &  &   & 0.615\\
     \hline
     Power-law profiles \\
     \hline
     GNFW & 1.94$\pm$0.51 & 4.51$\pm$0.61 & -0.09$\pm$0.08 & 7.79$\pm$0.90 & 1.23$\pm$0.18  & 0.414\\
     $\beta$ model &  & &  &  &   & 0.366\\
     Single Power-Law & 1.36$\pm$0.09 & 2.37$\pm$0.24 & 2.03$\pm$0.22 & 1.00 &   & 0.431\\
     Power-Law & 1.34$\pm$0.08 & 2.66$\pm$0.54 & 2.00 & 1.14$\pm$0.23 &  &  0.428\\
     Double Power-Law & 1.23$\pm$0.27 & 2.63$\pm$0.60 & 0.05$\pm$0.09 & 2.16$\pm$0.33 &  & 0.433\\
     \hline
     \end{tabular}
     \tablefoot{Values without uncertainties correspond to specific cases where one or more parameters were fixed during the fitting procedure.}

\end{table*}

\begin{table*}[h]
    \caption{Best-fit parameters of the observed galaxy overdensity profiles around filaments at $0.1 < z < 0.3$.}
        \label{tab:appendix2}
    \centering
    \begin{tabular}{ccccccc}
    \hline
     Profile & $\delta_{0}$ & $r_{f}$ (Mpc) & $\alpha$ & $\beta$ & $\gamma$ &  $\chi^{2}_{\nu}$ \\
     \hline
     Exponential profiles \\
    \hline
     Einasto & 1.03$\pm$0.076 & 6.79$\pm$0.61 & 1.14$\pm$0.15 &  &  & 0.2331\\
     Exponential & 1.12$\pm$0.04 & 6.10$\pm$0.37 & 1.00 &  &  & 0.2438\\
     Gaussian & 0.81$\pm$0.03 & 6.07$\pm$0.25 & 2.00&  &   & 0.3956\\
     \hline
     Power-law profiles \\
     \hline
     GNFW &  &  &  &  &  &  \\
     $\beta$ model &  &  &  &  &  &  \\
     Single Power-Law & 0.90$\pm$0.05 & 5.71$\pm$0.42 & 2.36$\pm$0.24 & 1.00 &   & 0.3501\\
     Power-Law & 0.89$\pm$0.04 & 9.08$\pm$2.29 & 2.00 & 1.90$\pm$0.66 &   & 0.3086\\
     Double Power-Law & 0.61$\pm$0.08 & 8.10$\pm$0.82 & 3.05$\pm$0.51 & 0.18$\pm$0.05 &  & 0.2994\\
     \hline
     \end{tabular}
     \tablefoot{Values without uncertainties correspond to specific cases where one or more parameters were fixed during the fitting procedure.}
\end{table*}

\begin{table*}[h]
    \caption{Best-fit parameters of the normalised galaxy overdensity profiles around filaments at $0.05 < z < 0.1$.}
        \label{tab:appendix3}
    \centering
    \begin{tabular}{ccccccc}
    \hline
     Profile & $\delta_{0}$ & $r_{f}$ & $\alpha$ & $\beta$& $\gamma$ &  $\chi^{2}_{\nu}$ \\
     \hline
     Exponential profiles \\
     \hline
     Einasto & 1.58$\pm$0.15 & 0.49$\pm$0.07 & 0.97$\pm$0.14 &  &  &  0.2234\\
     Exponential & 1.55$\pm$0.08 & 0.50$\pm$0.03 & 1.00 &  &  & 0.2222\\
     Gaussian & 1.17$\pm$0.05 & 0.46$\pm$0.02 & 2.00&  &   & 0.4114\\
     \hline
     Power-law profiles \\
     \hline
     GNFW & 1.76$\pm$0.82 & 0.98$\pm$0.24 & 3.80$\pm$0.47 & -2.88$\pm$0.34 & -1.24$\pm$0.34  & 0.2329\\
     $\beta$-model &  &  &  &  &   & \\
     Single Power-Law & 1.35$\pm$0.09 & 0.42$\pm$0.04 &  2.13$\pm$0.23 & 1.00 &  & 0.2591\\
     Power-Law &  &  &  &  &  &  \\
     Double Power-Law & 1.19$\pm$0.21 & 0.47$\pm$0.09 & 2.26$\pm$0.36 & 0.06$\pm$0.08 &  & 0.2572\\
     \hline
     \end{tabular}
      \tablefoot{Values without uncertainties correspond to specific cases where one or more parameters were fixed during the fitting procedure.}
\end{table*}

\begin{table*}[h]
    \caption{Best-fit parameters of the normalised galaxy overdensity profiles around filaments at $0.1 < z < 0.3$.}
        \label{tab:appendix4}
    \centering
    \begin{tabular}{ccccccc}
    \hline
     Profile & $\delta_{0}$ & $r_{f}$ & $\alpha$ & $\beta$ & $\gamma$ &  $\chi^{2}_{\nu}$ \\
     \hline
     Exponential profiles \\
     \hline
     Einasto & 1.06$\pm$0.07 & 0.90$\pm$0.08 & 1.12$\pm$0.13 &  &  & 0.0807\\
     Exponential & 1.13$\pm$0.05 & 0.83$\pm$0.05 & 1.00 &  &  & 0.0878\\
     Gaussian & 0.83$\pm$0.03 & 0.80$\pm$0.03 & 2.00&  &   & 0.2572\\
     \hline
     Power-law profiles \\
     \hline
     GNFW &  &  &  &  &  &   \\
     $\beta$-model &  &  &  &  &  &   \\
     Single Power-Law & 0.95$\pm$0.05 & 0.72$\pm$0.05 &  2.13$\pm$0.19 & 1.00 &  & 0.1475\\
     Power-Law & 0.94$\pm$0.04 & 0.93$\pm$0.18 & 2.00 & 1.37$\pm$0.32 &  &  0.2096\\
     Double Power-Law & 0.72$\pm$0.10 & 0.96$\pm$0.12 & 2.57$\pm$0.37 & 0.12$\pm$0.05 &  & 0.1252\\
     \hline
     \end{tabular}
     \tablefoot{Values without uncertainties correspond to specific cases where one or more parameters were fixed during the fitting procedure.}
\end{table*}

\end{appendix}
\end{document}